\documentclass{Interspeech2024}
\usepackage{amssymb}
\usepackage{pifont}
\usepackage[symbol]{footmisc}
\usepackage{multirow}

\interspeechcameraready

\title{RIR-SF: Room Impulse Response Based Spatial Feature for Target Speech Recognition in Multi-Channel Multi-Speaker Scenarios}

\name[affiliation={1,2*}]{Yiwen}{Shao}
\name[affiliation={2*}]{Shi-Xiong}{Zhang}
\name[affiliation={2}]{Dong}{Yu}
\address{
  $^1$Center for Language and Speech Processing, Johns Hopkins University, Baltimore, MD, USA\\
  $^2$Tencent AI Lab, Bellevue, WA, USA}
\email{yshao18@jhu.edu, zhangshixiong@gmail.com, DongYu@ieee.org}


\begin{document}

\maketitle
\renewcommand*{\thefootnote}{\fnsymbol{footnote}}
\footnote{This work was done while Yiwen was a research intern, Shi-Xiong was a researcher at Tencent AI Lab, USA. }
\begin{abstract}
Automatic speech recognition (ASR) on multi-talker recordings is challenging. Current methods using 3D spatial data from multi-channel audio and visual cues focus mainly on direct waves from the target speaker, overlooking reflection wave impacts, which hinders performance in reverberant environments. Our research introduces RIR-SF, a novel spatial feature based on room impulse response (RIR) that leverages the speaker's position, room acoustics, and reflection dynamics. RIR-SF significantly outperforms traditional 3D spatial features, showing superior theoretical and empirical performance. We also propose an optimized all-neural multi-channel ASR framework for RIR-SF, achieving a relative 21.3\% reduction in CER for target speaker ASR in multi-channel settings. RIR-SF enhances recognition accuracy and demonstrates robustness in high-reverberation scenarios, overcoming the limitations of previous methods.
\end{abstract}

\section{Introduction}

\label{sec:intro}
As speech techniques and deep neural networks have advanced, significant enhancements have been observed across various automatic speech recognition (ASR) benchmarks \cite{hinton2012deep, wang2019espresso, shao2020pychain, zhang2020pushing}. Nevertheless, the recognition of multi-channel multi-talker overlapped speech remains a formidable undertaking, primarily owing to the presence of interfering speakers and background noise \cite{haeb2020far, masuyama2023end}.

The conventional approach involves a pipeline-based paradigm, where an initial speech separation module separates the clean target speech for transcription by the backend ASR system \cite{yu2021audio, gu2022towards}. In contrast, a recent "All-in-one" system, explored in \cite{shao2022multi}, has shown promising results. This approach eliminates the need for explicit speech separation modules and doesn't rely on parallel clean separated speech as supervision during training, which is often unavailable for real data. Moreover, it offers advantages in terms of faster convergence and inference speed, making it particularly suitable for large-scale applications. 

With this objective in mind, having a high-quality and discriminative input feature becomes particularly crucial for this approach. To identify the target speech from the mixture,  additional information about the target speaker is required. Common clues such as speaker's voiceprint based features \cite{vzmolikova2019speakerbeam,wang2018voicefilter}, vision clues \cite{gu2020multi,ephrat2018looking, wu2019time} and location based features \cite{chen2018multi,wang2018combining,gu2019neural} have been shown to yield advantages with high accessibility and robustness. Among these features, the spatial feature, originally introduced in \cite{chen2018multi}, has gained widespread adoption, with its recent 3D version demonstrating state-of-the-art performance across various tasks, including speech separation and ASR \cite{gu2022towards, shao2022multi}. 

However, strong reverberation poses challenges for the 3D spatial feature due to its inability to account for reflection waves. Regular dereverberation, whether as preprocessing or part of an end-to-end neural network \cite{li2023audio}, is computationally intensive and time-consuming, hindering future development. As a result, instead of eliminating the reverberation, we propose to utilize the reverberation, that is, the room impulse response (RIR) of the target speaker to the microphone array as an extra input clue, to extract a new version of spatial feature for the downstream tasks which we named \textbf{RIR-SF}.

\noindent{\bf Contributions:} In this work, we present a novel spatial feature created by convolving the Room Impulse Response (RIR) corresponding to the target speaker with multi-talker speech. Additionally, we establish a dedicated neural network architecture for the extraction of this innovative spatial feature. The paper includes both theoretical analysis and experimental results to underscore its superior performance when compared to the conventional 3D SF.

\vspace{-0.2cm}
\section{Problem with 3D Spatial Feature}
\vspace{-0.1cm}
In this section, we will formulate the basic signal model in both time and frequency domain and then discuss the current problem of 3D Spatial Feature.
\subsection{Signal Model}

\textit{Time Domain:} With a $M$-channel microphone array, the noisy speech $Y^m(n)$ of the $m$-th channel can be formulated as:
\begin{align}
    Y^m(n) &= \sum_{i=1}^{I}\hat{X}^m_{i}(n) + e^m(n), \\
\hat{X}^m_{i}(n) &= R_i^m(n) * X_i(n) \label{eq:time_reverb}
\end{align}
where $*$ denotes convolution and $n$ is the time index. $\hat{X}^m_i(n)$ is the reverberant speech resulting from the $i$-th speech signal $X_i(n)$ convolved with its RIR $R_i^m(n)$ to the $m$-th channel, $e^m(n)$ is the ambient noise.

\textit{STFT Domain}: When expressing the signal model of Equation (\ref{eq:time_reverb}) in the STFT domain, because of the fact that the RIR's duration significantly exceeds the analysis window's length, the convolution in Equation \ref{eq:time_reverb} shifts from being a multiplication in the STFT domain to becoming a convolution with the inter-frame and inter-frequency \cite{avargel2007system}:
\begin{align}
    Y^m(t,f) &= \sum_{i=1}^{I} \hat{X}_i(t, f) + e^m(t, f) \\
    \hat{X}^m_{i}(t, f) &= \sum_{f'=0}^{F-1} R_i^m(t,f,f') * X_i(t, f)
    \label{eq:stft_reverb}
\end{align}
where $X_i(t, f)$ is the STFT coefficient of $i$-th source at frame index $t$ and frequency index $f$. $R_i^m(t,f,f')$ is a set of band-to-band $(f'= f)$ and crossband $(f' \neq f)$ filters derived from $R_i^m(n)$, and the convolution with $X_i(t, f)$ is applied to the time axis.

\begin{figure}\vspace{-0.1cm}
\centering
    \includegraphics[width=80mm, height=60mm]{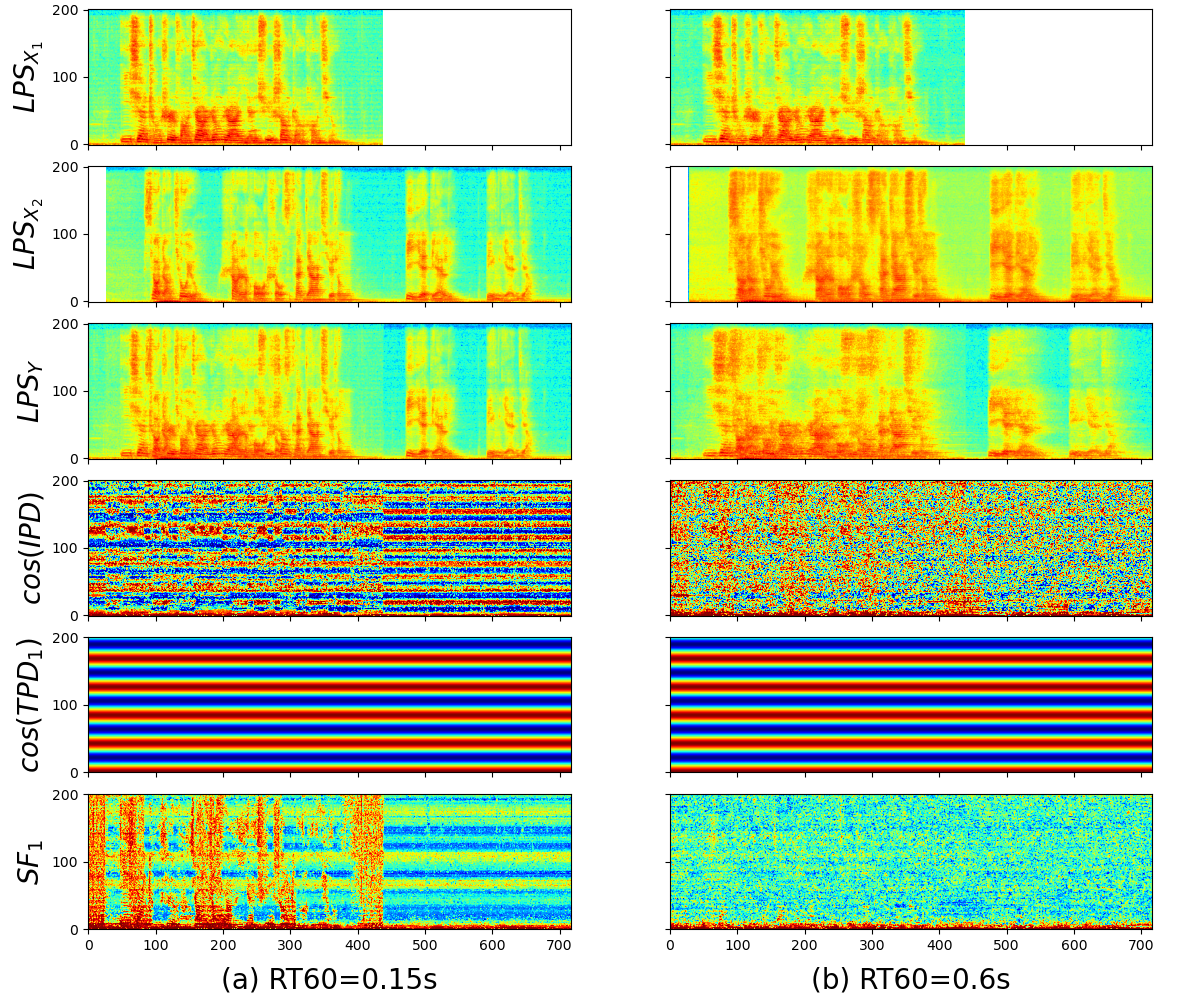}
    \caption{\small An illustration of 3D spatial features with (a) weak reverberation and (b) strong reverberation. In (a), $\mathbb{SF}_1$ matches well with the pattern of the Log Power Spectrum (LPS) of the target speaker $X_1$. While in (b), $\mathbb{SF}_1$ fails to identify the target source from the mixture.}\vspace{-0.4cm}
    \label{fig:3d_sf}
\end{figure}

While Equation \ref{eq:stft_reverb} is precise, it is seldom employed practically due to the high complexity associated with cross-band filtering. Alternatively, the \textbf{convolutive transfer function (CTF)} approximation \cite{talmon2009convolutive, li2019multichannel} can be adopted, which involves excluding the cross-band filters. And Equation \ref{eq:stft_reverb} can be re-formulated as:
\begin{equation}
    \hat{X}^m_{i}(t, f) = R_i^m(t,f) * X_i(t, f) \label{eq:ctf}
\end{equation}
where $R_i^m(t,f)$ is the STFT representation of $R_i^m(n)$.

A further simplification is \textbf{multiplicative transfer function (MTF)} approximation \cite{avargel2007multiplicative} that only considers the direct wave from the source to the microphone and ignores all reflection waves:
\begin{equation}
    \hat{X}^m_{i}(t, f) = R_i^m(0, f) X_i(t, f) \label{eq:mtf}
\end{equation}
where $R_i^m(0, f)$ is the STFT of the impulse response of the direct wave from source $i$ to microphone $m$. 
And in such way, $[R_i^0(n)(0, f), R_i^1(n)(0, f), ..., R_i^M(n)(0, f)]$ \textbf{forms the steering vector for source $i$.} 

\subsection{3D Spatial Feature from MTF/CTF Perspective}
\begin{figure}\vspace{-0.1cm}
\centering
    \vspace{-0.5cm}
    \includegraphics[width=80mm, height=60mm]{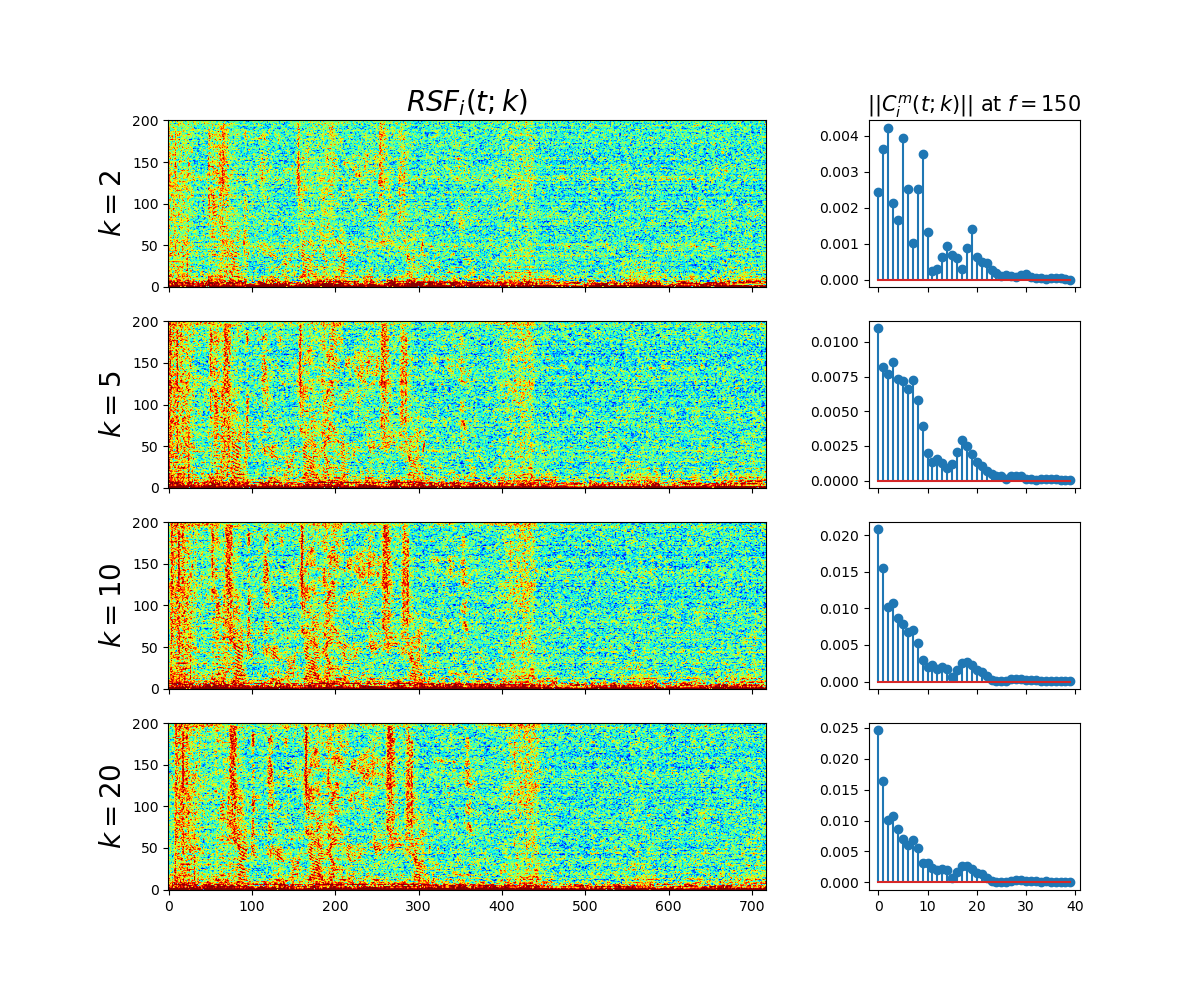}
    \vspace{-0.5cm}
    \caption{\small An illustration of $\mathbb{RSF}_i(t;k)$ and $\lVert C_i^{m}(t;k)\rVert$ for the example in Figure \ref{fig:3d_sf} (b) with different k.}\vspace{-0.4cm}
    \label{fig:rir_sf}
\end{figure}
3D Spatial Feature (3D-SF) \cite{shao2022multi} \cite{gu20213d}  is the state-of-the-art spatial cues used to indicate the dominance of the target source in the T-F bins. It can be formulated as the cosine similarity between the target-dependent phase difference (TPD) and the interchannel phase differences (IPD):
\begin{align}
    &\mathbb{SF}_i(t) = \cos (\text{IPD}^{(m)}(t) - \text{TPD}_i^{(m)}(t)) \\
    &\text{IPD}^{(m)}(t) = \angle Y^{m_1}(t) - \angle Y^{m_2}(t) \label{eq:ipd}\\
    &\text{TPD}^{(m)}_i(t) = \angle R_i^{m_1}(0) - \angle R_i^{m_2}(0)
\end{align}
where $\angle$ is the phase of a complex number, $(m)$ is a selected microphone pair $(m_1, m_2)$. If multiple pairs of microphones are chosen, the resultant $\mathbb{SF}_i(t)$ would be the sum of them. \textbf{Please note that in the above equations, we omit the frequency index $f$ as all the formulations are on a per-frequency basis}. From now on, unless explicitly stated, we will keep the same notation throughout the paper. 

Suppose the T-F bin of the mixed speech $Y^m(t)$ is dominated by source $i$, (i.e. $Y^m(t) = \hat{X}_i^m(t)$), and take the MTF approximation Equation \ref{eq:mtf} into Equation \ref{eq:ipd}, we have:
\begin{equation}
\begin{aligned}
    \text{IPD}^{(m)}(t) = &\angle \hat{X}_i^{m_1}(t) - \angle \hat{X}_i^{m_2}(t) \\
    =& \angle R_i^{m_1}(0) X_i(t) - \angle R_i^{m_2}(0) X_i(t) \\
    =& (\angle R_i^{m_1}(0) + \angle X_i(t)) - (\angle R_i^{m_2}(0)\\
     &+ \angle X_i(t)) \\
     =& \angle R_i^{m_1}(0) - \angle R_i^{m_2}(0) \\
     =& \text{TPD}^{(m)}_i(t) \label{eq:sf_mtf}
\end{aligned}
\end{equation}
In this way, $\mathbb{SF}_i(t) = \cos (\text{IPD}^{(m)}(t) - \text{TPD}_i^{(m)}(t)) = 1$ and thus serves as a strong indicator for the dominance of target source $i$ in the T-F bin.

However, when the reverberation is severe and unignorable, which for example, in a typical living room environment with RT60 ranges from 0.3s to 0.7s, MTF is no longer sufficient to model the signal while CTF is more appropriate. Take CTF approximation Equation \ref{eq:ctf} into Equation \ref{eq:ipd}, we have:
\begin{equation}
    \begin{aligned}
       & \text{IPD}^{(m)}(t) = \angle R_i^{m_1}(t) * X_i(t) - \angle R_i^{m_2}(t) * X_i(t) \\
        &= \angle (\sum_{k=0}^K R^{m_1}_i(k)X_i(t+k)) - \angle (\sum_{k=0}^K R^{m_2}_i(k)X_i(t+k)) \\
        &\neq  \text{TPD}^{(m)}_i(t) \label{eq:sf_ctf}
    \end{aligned}
\end{equation}

\begin{figure*}[t]
  \hspace{-0.5cm}
  \includegraphics[width=1.0\linewidth]{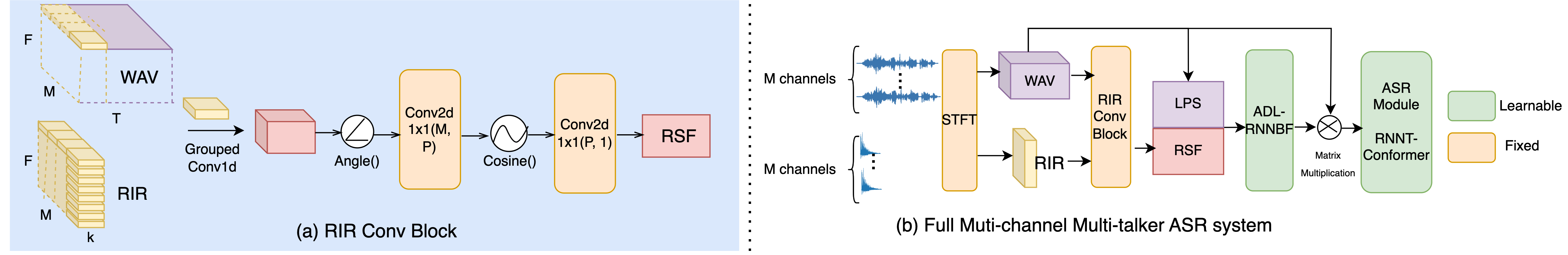}%\vspace{-0.1cm}
  \caption{\small (a) RIR Conv Block that utilizes convolution layers to extract RIR-based spatial feature $\mathbb{RSF}_i(t;k)$; (b) The whole Multi-channel Multi-speaker ASR system that combines $\mathbb{RSF}_i(t;k)$ and ADL-RNNBF.} %\vspace{-0.2cm}
  \label{fig:diagram}
\end{figure*}

Consequently, $\mathbb{SF}_i(t)$ wouldn't be close to 1 even if the mixture T-F bin is dominated by the target speaker $i$ and then become less informative and discriminative for the downstream tasks. Figure \ref{fig:3d_sf} shows a comparison between $\mathbb{SF}_i(t)$ under 2 scenarios where weak (RT60=0.15s) and strong (RT60=0.6s) reverberation exists respectively. 

\section{RIR Based Spatial Feature}
We propose to solve the problem by utilizing the spatial information contained in the RIR. We firstly convolve the mixture speech $Y^{m}(t)$ with the \textbf{conjugate} of the first $k$ frames of $R_i^{m}(t)$ and then extract its phase, which we define as RIR-convolved Phase (RP): 
\begin{equation}
    \text{RP}_i^{m}(t;k) = \angle Y^{m}(t) * R_i^{m}(t)^H |_{t=0:k-1}
\end{equation}   
$k$ serves as a hyper-parameter for this term. 

Then we define a new spatial feature \textbf{RIR-SF}: $\mathbb{RSF}_i(t;k)$ as the cosine of RP difference between a pair of channels:
\begin{equation}
        \mathbb{RSF}_i(t;k) := \cos (\text{RP}_i^{m_1}(t;k) - \text{RP}_i^{m_2}(t;k))
\end{equation}
\textbf{Specially, when $k=1$, i.e. only considering the direct wave, $\mathbb{RSF}_i(t;k)$ can be simplified to $\mathbb{SF}_i(t)$}:
\begin{equation}
\begin{aligned}
        &\mathbb{RSF}_i(t;k=1) = cos (\angle Y^{m_1}(t) R_i^{m_1}(0)^H - \angle Y^{m_2}(t) R_i^{m_2}(0)^H ) \\
        &= cos (\angle Y^{m_1}(t) - \angle R_i^{m_1}(0) - \angle Y^{m_2}(t) + \angle R_i^{m_2}(0)) \\
        &= cos (\text{IPD}^{(m)}(t) - \text{TPD}_i^{(m)}(t)) \\
        &= \mathbb{SF}_i(t)
\end{aligned}
\end{equation}
\textbf{In other words, $\mathbb{RSF}_i$ is an extension to the previous $\mathbb{SF}_i$}.
Next, we will demonstrate how this novel suggested spatial feature can effectively highlight the prominence of the target source in TF bins, even in the presence of substantial reverberation.

Suppose in the selected T-F bin, the target speaker $i$ is dominating the mixture, i.e. $Y^{m}(t) = \hat{X}_i^m(t) = R_i^m(t) * X_i(t)$, we can write $\text{RP}_i^{m}(t;k)$ as:
\begin{equation}
    \begin{aligned}
       \text{RP}_i^{m}(t;k) &= \angle R_i^{m}(t) * X_i(t) * R_i^{m}(t)^H |_{t=0:k-1} \\
        &= \angle X_i(t) * (R_i^{m}(t) * R_i^{m}(t)^H |_{t=0:k-1}) \label{eq:rp}
    \end{aligned}
\end{equation}
In Equation \ref{eq:rp}, $\text{RP}_i^{m}(t;k)$ becomes the phase of the clean anechoic speech $X_i(t)$ convolved with a new kernel $C_i^{m}(t;k)$, where
\begin{equation}
    \begin{aligned}
       C_i^{m}(t;k) &= R_i^{m}(t) * R_i^{m}(t)^H |_{t=0:k-1} \\
        &= \sum_{n=0}^{k-1} R_i^{m}(t+n) R_i^{m}(n)^H \\
    \end{aligned}
\end{equation}
Specially, we have the first element of $C_i^{m}(t;k)$ as
\begin{equation}
    \begin{aligned}
    C_i^{m}(0;k) &= \sum_{n=0}^{k-1} R_i^{m}(n) R_i^{m}(n)^H \\
    &= \sum_{n=0}^{k-1} \lVert R_i^{m}(n) \rVert_2
    \end{aligned}
\end{equation}
where $C_i^{m}(0;k) \in \mathbb{R}$. Taking $C_i^{m}(t;k)$ back into Equation \ref{eq:rp} we have
\begin{equation}
    \begin{aligned}
        \text{RP}_i^{m}(t;k) &= \angle X_i(t) * C_i^{m}(t;k) \\
        &= \angle \sum_{n=0}^{K-1} X_i(t+n) C_i^{m}(n; k) \\
        & = \angle (X_i(t)C_i^{m}(0;k) + \sum_{n=1}^{K-1} X_i(t+n) C_i^{m}(n; k)) \\ 
        &\approx \angle (X_i(t)C_i^{m}(0;k)) = \angle X_i(t) \label{eq:rp_2}
    \end{aligned}
\end{equation}
where $K$ is the total length of $C_i^m(t;k)$. The approximation in Equation \ref{eq:rp_2} is valid when $\lVert C_i^{m}(0;k)\rVert \gg \lVert C_i^{m}(t;k)\rVert$ for all $t > 0$. Fortunately, due to the fact that RIR is a fast decaying function, this condition can be easily satisfied when $k$ is sufficiently large, as shown in Figure \ref{fig:rir_sf}. In this way, $\text{RP}_i^{m}(t;k)$ becomes totally channel independent, which leads to a desiring result of $\mathbb{RSF}_i(t;k) = 1$ when the target speaker $i$ dominates the mixture TF bin.

%\noindent \textbf{Previous 3D feature is a special case of RIR-SF.}
\subsection{Why RIR-SF is better}
As has been pointed out, the previous 3D spatial feature is a special case of $\mathbb{RSF}(t;k)$ with $k=1$. From both Figure \ref{fig:rir_sf} and Equation (\ref{eq:rp_2}) we know that the sharpness of $\lVert C_i^{m}(t;k)\rVert$ determines the quality of $\mathbb{RSF}(t;k)$. By increasing $k$ to a reasonable value, we can cancel out the impact of reflection waves in the phase computation.

\section{System}
\subsection{RIR Conv Block}
%\vspace{-0.2cm}
We build a \texttt{RIR Conv Block} (Figure \ref{fig:diagram} (a)) to extract the proposed $\mathbb{RSF}_i(t;k)$ in a fully neural way. As discussed in section 2 and 3, the RIR convolution to the multi-channel speech is applied to the time axis and is on both per-frequency and per-channel basis. We break the conjugate of the target RIR $R_i^H \in C^{[M \times F \times k]}$ into $M \times F$ groups of $(1 \times k)$ kernels, and convolve each of them with the corresponding split of $Y \in C^{[1\times T]}$ separately, resulting in an intermediate complex output of shape $[M\times F\times T]$. After that, we obtain $\text{RP}_i^{m}(t;k)$ and $\mathbb{RSF}_i(t;k)$ through 2 specially initialized \texttt{Conv2d\_1x1} layers, which is used for doing pair-wise difference and summation respectively. Please note that we keep these 2 layers fixed in this work to follow the strict formulation of the proposed $\mathbb{RSF}_i(t;k)$. More work may be done in the future to explore the potential of making the RIR convolution fully learnable.

\vspace{-0.2cm}
\subsection{ADL-RNNBF}
\vspace{-0.2cm}
The state-of-the-art neural beamformer ADL-RNNBF \cite{zhang2021adl} is found to be consistently beneficial for multi-channel multi-talker ASR and separation \cite{shao2022multi,gu2022towards}, especially when the spatial feature itself is not sufficiently discriminative. For this reason, we keep it an optional enhancement module to test its interactivity with the proposed RIR-based spatial feature. It has the same structure and hyper-parameter as in \cite{zhang2021adl}, except that we used concatenated LPS and $\mathbb{RSF}_i(t;k)$ as input and didn't apply the inverse STFT to transfer its output back to the waveform.

\vspace{-0.2cm}
\subsection{RNNT-Conformer}
\vspace{-0.2cm}
A pruned version of RNNT model \cite{kuang2022pruned} is used as the ASR backbone.  The network is formed by a 12-layer 4-head Conformer \cite{gulati2020conformer} encoder with 512 attention dimensions and 2048 feed-forward dimensions.

%\vspace{-0.2cm}
\section{Experiments}
%\vspace{-0.2cm}
\subsection{Dataset}
%\vspace{-0.2cm}
%\subsection{Data}
We simulate a multi-channel reverberant dataset based on AISHELL-1 \cite{bu2017aishell} for experiments. \textbf{Pyroomacoustics \cite{scheibler2018pyroomacoustics} with image-source method (ISM)\cite{lehmann2008prediction} is used as the toolkit for RIR generation and estimation, with \textit{(room size, RT60, mic position, speaker positions)} as parameters}. The room size is ranging from [3, 3, 2.5] to [8, 6, 4] meters, with one microphone array and 2 speakers in each room. An 8-element non-uniform linear array with spacing 15-10-5-20-5-10-15 cm is fixed, with random position inside the room. We have 2 settings for RT60, ranging from [0.1, 0.6] seconds and [0.5, 0.7] seconds, which corresponds to \textbf{weak} and \textbf{strong} reverberation scenarios respectively. Thus, 100,000 sets of RIR are generated offline for each setting. Note that only the weak reverberation is used for training as we've found simply training with stronger reverberation doesn't improve system performance generally.  Finally, the multi-channel reverberant overlapped speech are synthesized on-the-fly during training by convolving the generated RIR with dry clean speech from the corresponding AISHELL split, with the signal-to-interference rate (SIR) randomly sampled from -6 to 6 dB. The overlap ratio of two speakers inside a mixed utterance are from 0.5 to 1. 

To test our proposed spatial feature's robustness to the estimated target RIR, we design 3 different RIR estimation situations as shown in Table \ref{tab:rir}. It should be noted that, with the relative position of microphones and the target speaker fixed in all 3 situations, the oracle 3D spatial information to $\mathbb{SF}$ is actually guaranteed.

\begin{table}[]
\caption{\small 3 different scenarios for target RIR estimation using image-source method (ISM). In sec1 and sec2, RT60 is randomly sampled from [0.3, 0.8] seconds. In sec2, room size, mic position and speaker position have a random shift from the ground truth within 0.5 meters in all 3 dimensions.}
\resizebox{1.0\columnwidth}{!}{%
\begin{tabular}{l|lllll}
\hline
\multicolumn{1}{c|}{\multirow{3}{*}{\textbf{scenario}}} &
  \multicolumn{5}{c}{\textbf{information obtained from vision input}} \\ \cline{2-6} 
\multicolumn{1}{c|}{} &
  \multicolumn{1}{c|}{\begin{tabular}[c]{@{}c@{}}room \\ size \end{tabular}} &
  \multicolumn{1}{c|}{\begin{tabular}[c]{@{}c@{}}mic \\ position \end{tabular}} &
  \multicolumn{1}{c|}{\begin{tabular}[c]{@{}c@{}}speaker \\ position \end{tabular}} &
  \multicolumn{1}{c|}{\begin{tabular}[c]{@{}c@{}}spk relative \\ pos to mic\end{tabular}} &
  \multicolumn{1}{c}{\begin{tabular}[c]{@{}c@{}} reflect factor \\ of wall (RT60)\end{tabular}} \\ \hline
ideal &
  \multicolumn{1}{l|}{\checkmark} &
  \multicolumn{1}{l|}{\checkmark} &
  \multicolumn{1}{l|}{\checkmark} &
  \multicolumn{1}{l|}{\checkmark} &
  \checkmark \\ \hline
sce1 &
  \multicolumn{1}{l|}{\checkmark} &
  \multicolumn{1}{l|}{\checkmark} &
  \multicolumn{1}{l|}{\checkmark} &
  \multicolumn{1}{l|}{\checkmark} &
  \ding{55} \\ \hline
sce2 &
  \multicolumn{1}{l|}{\ding{55}} &
  \multicolumn{1}{l|}{\ding{55}} &
  \multicolumn{1}{l|}{\ding{55}} &
  \multicolumn{1}{l|}{\checkmark} &
  \ding{55} \\ \hline
\end{tabular}
}
\label{tab:rir}
\end{table}

\begin{table}[]
\caption{\small CER \% on simulated AISHELL-1 dev/test set with weak (RT60=[0.1, 0.6]s) and strong (RT60=[0.5, 0.7]s) reverberation.}
\resizebox{1.0\columnwidth}{!}{%
\begin{tabular}{l|l|l|l|l|l}
\hline \textbf{scenario} & \textbf{feature} & \textbf{BF} & $\mathrm{k}$ & \textbf{weak} & \textbf{strong} \\
\hline single-spk & - & \ding{55} & - & $8.57 / 9.71$ & $10.64 / 11.99$ \\
 \hline ideal & $\mathbb{SF}$ & \ding{55} & - & $14.11 / 15.67$ & $20.28 / 21.26$ \\
ideal & $\mathbb{RSF}$ & \ding{55} & 2 & $12.58 / 14.21$ & $17.58 / 19.33$ \\
 ideal & $\mathbb{RSF}$ & \ding{55} & 10 & \textbf{10.4 / 11.75} & \textbf{11.37 / 12.71} \\
 ideal & $\mathbb{RSF}$ & \ding{55} & 20 & $10.49 / 11.9$ & $11.33 / 13.03$ \\
\hline ideal & $\mathbb{SF}$ & \checkmark & - & \textit{11.93 / 13.43} & \textit{15.48 / 17.43} \\
 ideal & $\mathbb{RSF}$ & \checkmark & 10 & \textbf{9.84 / 11.06} & \textbf{11.13 / 12.66} \\
\hline sce1 & $\mathbb{RSF}$ & \ding{55} & 10 & $11.16 / 12.87$ & $11.57 / 13.26$ \\
 sce1 & $\mathbb{RSF}$ & \checkmark & 10 & \textbf{9.78 / 11.07} & \textbf{10.87 / 12.3} \\
\hline sce2 & $\mathbb{RSF}$ & \ding{55} & 10 & $22.37 / 24.33$ & $22.49 / 24.76$ \\
 sce2 & $\mathbb{RSF}$ & \checkmark & 10 & \textbf{12.18 / 14.31} & \textbf{12.19 / 14.4} \\
\hline
\end{tabular}}
\label{tab:cer}
\end{table}
%%%%%%%%%%%%%%%%%%%%%%%%%%%%%%%%%%%%%%%%%%%%

\subsection{Experimental Results Analysis}

Table \ref{tab:cer} shows a comprehensive comparison between the proposed $\mathbb{RSF}$ and the previous state-of-the-art 3D $\mathbb{SF}$ under different situations and modules. Note that unlike other numbers shown in Table \ref{tab:cer}, the first row is both trained and tested on non-overlapped single-speaker speech, which is presented here as an ASR upper bound for the following systems to be compared with.

\vspace{0.1cm}
\noindent\begin{minipage}{\linewidth}
\begin{itemize}[leftmargin=*, noitemsep, topsep=0pt]
    \item \textbf{weak vs strong reverberation:} The experiments on the $\mathbb{SF}$ corroborate our previous discussion in section 2.2 that strong reverberation will significantly degrade its performance on multi-speaker ASR performance (-6.17\%/-5.59\%), even with the enhancement from ADL-RNNBF (-3.55\%/-4.00\%), which are still much higher than an expected performance drop on the single-speaker speech (-2.07\%/-2.28\%). However, with the proposed $\mathbb{RSF}$, when using RIR with $k \geq 0.1 s$, the performance gap between weak and strong reverberation is always within a reasonable range (1\% to 2\%), showing its robustness against reverberation.
    \item \textbf{Robustness to RIR estimation:} When provided with ground truth spatial information to estimate the target RIR, $\mathbb{RSF}$ achieves highly satisfactory results, whether with or without ADL-RNNBF. In reality, however, we can not always expect to have all these spatial information precise, especially for the virtual parameter of RT60 which can not be measured. When we only have RT60 wrong, $\mathbb{RSF}$ is almost unaffected, as can be shown from comparison between ideal scenario and sce1. But when we also get wrong information on other spatial information too, as illustrated by sce2, the enhancement from ADL-RNNBF becomes indispensable. And in such worst case scenarios, the proposed $\mathbb{RSF}$ can still obtain comparable results to the previous state-of-the-art system with 3D spatial feature under weak reverberation (-1.02\%/-1.44\%) but much better results under strong reverberation (3.29\% /3.03\%).
\end{itemize} 
\end{minipage}

\section{Conclusions}
%\vspace{-0.2cm}
% In summary, this study delved into the theoretical underpinnings of spatial features, leading to the development of a novel variant known as RIR-SF. Our innovative approach involves convolving multi-talker speech with the target Room Impulse Response (RIR) to generate this distinctive feature. Furthermore, we devised a dedicated neural system for the extraction and application of RIR-SF in downstream tasks, resulting in significant performance gains over the established 3D spatial feature, thereby demonstrating its superiority for the multi-channel multi-talker ASR task. We believe RIR-SF can achieve similar improvement on other tasks such as speech separation and enhancement as well.

In conclusion, this research delves deep into the nuances of spatial features, culminating in the inception of the innovative RIR-SF. By convolving multi-talker speech with the targeted Room Impulse Response (RIR), we've forged this unique feature. Additionally, we've architected an all-neural multi-channel system dedicated for the extraction and utilization of RIR-SF in downstream tasks. Our results reveal marked improvements over the prevailing 3D spatial feature, thereby demonstrating its superiority for the multi-channel multi-talker ASR task. Given its capabilities, we are optimistic about RIR-SF enhancing performance in related domains such as speech separation and enhancement.

\bibliographystyle{IEEEtran}
\bibliography{mybib}

\end{document}